# Site adaptation with machine learning for a Northern Europe gridded solar radiation product

Sebastian Zainali[1,*], Dazhi Yang[2], Tomas Landelius[3], Pietro E. Campana[1,*]

1 Mälardalen University, Department of Sustainable Energy Systems, Västerås SE 72123, Sweden

2. School of Electrical Engineering and Automation, Harbin Institute of Technology, Harbin, Heilongjiang, China

3. Swedish Meteorological and Hydrological Institute, Norrköping SE 601 76, Sweden

- Corresponding author:   sebastian.zainali@mdu.se, pietro.campana@mdu.se

## Abstract

Gridded global horizontal irradiance (GHI) databases are fundamental for analysing solar energy applications' technical and economic aspects, particularly photovoltaic applications. Today, there exist numerous gridded GHI databases whose quality has been thoroughly validated against ground-based irradiance measurements. Nonetheless, databases that generate data at latitudes above 65˚ are few, and those available gridded irradiance products, which are either reanalysis or based on polar orbiters, such as ERA5, COSMO-REA6, or CM SAF CLARA-A2, generally have lower quality or a coarser time resolution than those gridded irradiance products based on geostationary satellites. Among the high-latitude gridded GHI databases, the STRÅNG model developed by the Swedish Meteorological and Hydrological Institute (SMHI) is likely the most accurate one, providing data across Sweden. To further enhance the product quality, the calibration technique called "site adaptation" is herein used to improve the STRÅNG dataset, which seeks to adjust a long period of low-quality gridded irradiance estimates based on a short period of high-quality irradiance measurements. This study, differing from the conventional statistical approaches, adopts machine learning for site adaptation. Nine machine-learning algorithms have been analysed and compared with conventional statistical ones to identify Sweden's most favourable technique for site adaptation. Three weather stations of SMHI are used for training and validation. The results show that, due to the spatio-temporal heterogeneity in model performance, no universal model can be identified, which suggests that site adaptation is a location-dependent procedure.

# 1 Introduction

Global energy consumption increased from ~28 TWh to ~173 TWh between 1950–2019 (Ritchie & Roser, 2020a). Until 2019, 78% of the consumption was from fossil fuels (Ritchie & Roser, 2020b), which has hitherto been the main driver of climate change, one of the most challenging problems concerning humanity (Lu et al., 2020). To mitigate climate change, developing appropriate and supporting policies and regulations on deploying renewable energy is thought to be the foremost step (Moosavian et al., 2013; IEA, 2021). The European Union (EU) has set progressive targets to reduce greenhouse gas emissions through 2050 (European Commission, 2016). The energy sector is responsible for 75% of the European Union's greenhouse gas emissions, and to ensure that the EU achieves the ambitious climate targets set, a revised Climate and Energy package was released in July 2021 (European Commission, 2021). The target is to reduce at least 55% of greenhouse gas emissions by 2030 compared to 1990, which demands significantly higher shares of renewable energy in the energy mix. In this regard, the current EU target of having at least 32% renewable energy by 2030 is insufficient, and according to the climate target plan, it needs to be adjusted to 38%-40% to broaden the prospect of reaching the greenhouse gas emissions target by 2030 (European Commission, 2020).

Sweden is a member of the EU and has a political agreement to have a 100% renewable electricity system by 2040 (Riksdagsförvaltningen, 2018). To achieve this national goal, the Swedish government has supported, among other renewables, the solar energy sector with subsidies for photovoltaic (PV) installations. A higher level of subsidy has increased the margin of profitability of PV installations. The cumulative grid-connected PV capacity has boosted from 4 MW in 2005 to 1226 MW in 2020 (Masson & Kaizuka, 2021). In 2020, the Swedish government decided to support the PV sector by giving a tax deduction on labour and materials with a roof of 50 kSEK instead of subsidies, and the policy commenced on January 1, 2021 (Energimyndigheten, 2021).

A means to support the solar PV sector is to develop gridded solar irradiance products at high spatio-temporal resolutions so as to better support siting, sizing, design, performance evaluation, and operation & management of PV systems. Campana et al. (2020) analysed several gridded global horizontal irradiance (GHI) databases and determined that the CM SAF SARAH-2 has the best performance among the five databases. The CM SAF SARAH-2 covers Africa, the Atlantic, Europe, and part of South America with a resolution of 0.05° × 0.05° and is based on the geostationary METEOSAT satellites (EUMETSAT CM SAF, 2021). However, a drawback of using CM SAF SARAH-2 is that the database does not provide GHI estimates for latitudes above 65°, owing to the limited field of view. On the other hand, the product based on the STRÅNG model, which attained the second-best result during the comparison (Campana et al., 2020), is not restricted by the latitude limit since STRÅNG also integrates data from polar orbiters into its modelling. Specifically, STRÅNG is developed by the atmospheric remote sensing group at the Swedish Meteorological and Hydrological Institute (SMHI). It outputs estimates of GHI, beam horizontal radiation, photosynthetic active radiation, and CIE-weighted UV irradiance (i.e., the irradiance of each wavelength in the UV is weighted by the weighting factor given by the Commision Internationale de l'Éclairage (CIE)-action spectrum that gives the CIE-weighted spectral irradiance). It has a temporal resolution of 1 h, covering the Nordic countries with a spatial resolution of 2.5 km × 2.5 km (Landelius et al., 2003; SMHI, 2021).

Solar resource assessment is the foremost step for developing a solar projection. However, owing to the high cost and time constraints, long-term ground-based radiometry measurement at the target site is rarely available. For that reason, one has to resort to performing resource assessments based on satellite-derived irradiance (Polo et al., 2016). Physical and empirical models are the two main approaches for estimating satellite-derived irradiance (Miller et al., 2013). Physical models are based on (reduced forms of) radiative transfer, which seeks to estimate the attenuating effect of the atmosphere on incoming radiation. The empirical models, on the other hand, regress the ground observations onto various predictors, such as the satellite visible channel's recorded intensity, such that when new predictors are available, the fitted regression can issue predictions accordingly.

Polo et al. (2020) have conducted a benchmarking study on several site-adaptation techniques to assess the quality improvement brought by those techniques on ten gridded irradiance products covering satellite-derived and reanalysis solar radiation data. It was found that most techniques can significantly improve at most sites regarding bias reduction. In parallel, some sites with high-quality satellite-derived irradiance did not get any noticeable improvement after site adaptation. To that end, the author emphasised that no universal procedure can apply to all possible combinations of sites and modelled datasets, and the quality improvement is, in the main, heterogeneous. Two commonly used site adaption techniques to reduce bias and improve the model performance in a given geographical area are linear regression and quantile mapping (Polo et al., 2016; Yang & Gueymard, 2021).

Since the scientific principle of site adaptation is one of regression, one needs not to restrict to statistical methods as Polo et al. (2016) did. Narvaez et al. (2021) have already used machine learning as a site-adaptation technique and showed improvements over the traditional quantile mapping. The advantage of using machine learning lies in its diversity and versatility, which have, on many occasions, been shown to be advantageous. There is a rich opportunity for developing machine and deep learning models for site adaptation. Notwithstanding, as is the case for statistical site adaptation, the performance of machine-learning-based site-adaptation models also depends highly on the local weather and irradiance regime, which is hard to know *a priori*. In other words, one cannot infer model performance just from experience with high certainty. Instead, several machine-learning models must be developed and compared to find the optimal model, according to specific performance criteria and under the local regime.

In this study, machine learning, given its ability to perform regression, is used as a site-adaptation strategy to improve the quality of the GHI generated by STRÅNG. Site adaptation, in solar energy meteorology, refers to correcting the bias in a long period of gridded data using a short period of ground-based measurements (Polo et al., 2016, 2020). In a large pool of available machine-learning models, several algorithms with distinct prediction mechanisms are chosen, which include support vector regression (SVR), *k*-nearest neighbour (k-NN), Bagging, LSBoost, XGBoost, CatBoost, artificial neural network (ANN), convolutional neural network (CNN), and long short-term memory (LSTM). The performance of the site-adapted GHI using these machine-learning techniques is compared to the GHI estimates from STRÅNG and two traditional statistical site-adaptation approaches (linear and quantile mapping), using ground-based measurements at three locations in Sweden.

## 2  Method

This section describes the methodology used to improve the STRÅNG database through site adaption and the selected machine learning algorithms. Section 2.1 presents the data used in this study. Section 2.2 describes the machine learning models employed in this study. Section 2.3 describes the hyperparameters optimisation used to increase the performance of the machine-learning models. Section 2.4 describes the error metrics used to analyse the machine-learning model estimations. Section 2.5 describes the sensitivity analysis conducted in this study.

## 2.1 Data

In Sweden, SMHI maintains several weather stations that perform solar irradiance measurements. In this work, site adaption is conducted at three locations: Kiruna, Norrköping, and Visby, as also listed in Table 1 and mapped in Fig. 1. These locations represent the extremes in terms of the availability of direct and diffuse irradiance in Sweden. On an annual basis, Kiruna has the highest amount of diffuse irradiance, while Visby has the highest amount of direct irradiance. Norrköping, on the other hand, may be representative of the Swedish territory as it does not have any of these extremes (Campana et al., 2020).

As for the gridded product, STRÅNG is a mesoscale model for solar radiation products. The resolution and geographical extent of STRÅNG have changed over the years. Between January 1999 and May 2006, the horizontal resolution was about 22 km × 22 km. Then it was increased to 11 km × 11 km since. Furthermore, starting from 2017, the product is at the current resolution of 2.5 km × 2.5 km. The input and output fields that are produced from the model are adapted to the mesoscale analysis system MESAN at SMHI. Input data also come from numerical weather predictions from MEPS (*SMHI*, 2021).

Table 1: Swedish Meteorological and Hydrological Institute weather stations under consideration.

| Locations  | Latitude (°N) | Longitude (°E) | Altitude (m) |
|------------|---------------|----------------|--------------|
| Kiruna     | 67.84         | 20.41          | 424          |
| Norrköping | 58.58         | 16.14          | 43           |
| Visby      | 58.67         | 18.34          | 49           |

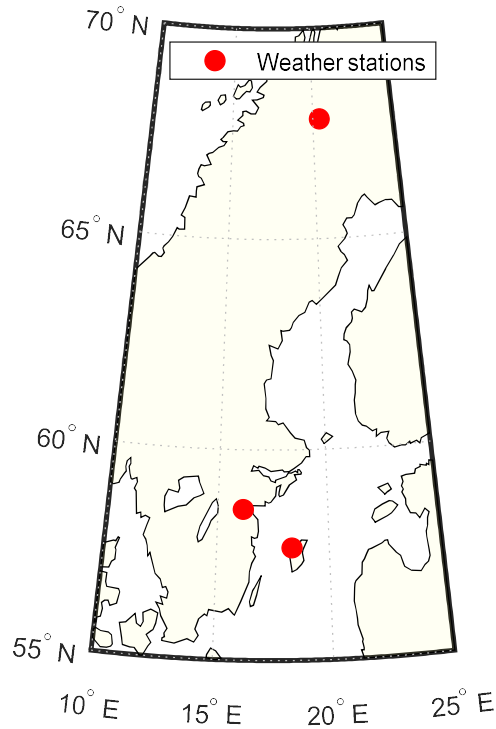

Fig. 1 Geographical locations of the three Swedish Meteorological and Hydrological Institute weather stations used in this study.

The weather station GHI measurement is used as the output target for the machine learning models during training, validation and testing. The weather stations in Kiruna, Norrköping, and Visby measure global irradiance with Kipp & Zonen CM21 pyranometer. SMHI does quality assurance routines based on routines developed for the Baseline Surface Radiation Network (BSRN) and are adjusted to fit the measurement program at the Swedish stations and tuned to the climatological limits in Sweden (Carlund, 2011). Traditional statistics approaches are evaluated with machine learning models to see their performance difference. The statistical approach uses the STRÅNG data as input and GHI measurements from the weather stations. The training dataset consists of data from 2008–2009, and the testing dataset consist of data from 2010–2015.

### 2.2 Machine-learning models for site adaptation

The predictors used for developing machine-learning models are STRÅNG GHI, solar azimuth, solar elevation, and time including months, days, and hours of the day. GHI from the weather station is used as the predicted parameter for the models. The machine learning algorithms used in this study can be classified into three categories: Shallow Models, Ensembles, and Deep Models. A summary is provided in Table 2.

Table 2: Machine learning methods.

| Shallow Models | Ensembles | Deep Models |
|---|---|---|
| ANN | Bagging | CNN |
| k-NN | LSBoost | LSTM |
| SVR | XGBoost | |
| | CatBoost | |

2.2.1   Shallow models

The k-NN method is used for classifying objects based on the closest training examples. The k-NN method can be used for regression (Carrera & Kim, 2020; Xu et al., 2021). However, the k-NN looks into the history to find cases that have the closest pattern to this case instead of using a learning base. Linear regressors are commonly used to minimise the sum of squared errors (Hastie et al., 2009). Several extensions of the linear regression are known as lasso, ridge, and elastic net with additional penalty parameters to minimise the complexity or reduce the number of features used in the model (Hastie et al., 2009).

In some cases, we only want to reduce the error to a certain degree between an acceptable range. In those cases, the SVR gives the flexibility to define an acceptable error in our model and find an appropriate line to fit the data (Carrera & Kim, 2020). A neural network is commonly used for predictive modelling, adaptive control, and applications to be trained via a dataset. The ANN modelling has become popular in the last decade due to being successful in several fields of medicine, mathematics, engineering, meteorology, and many other subjects (Şenkal & Kuleli, 2009).

2.2.2   Ensembles

Ensemble methods combine the strengths of a set of models into one predictive model where several ensemble learners exist to decrease the variance, bias, or improve the predictions (Hastie et al., 2009). Several ensemble methods construct regression trees when predicting the outcome for the given regression problem. The purpose of Bagging is to decrease the model's bias by averaging the prediction over multiple estimates. Boosting methods decrease the model's bias by training different models sequentially to improve the previous models generated (Carrera & Kim, 2020). The difference between boosting and bagging methods is that boosting learners are trained on a weighted data version. Extreme gradient boosting (XGBoost) is an improved gradient boosting technique that has added several parameters to improve the algorithm's prediction accuracy (Choi, 2019). The XGBoost uses Newton's tree boosting to optimise the learning of tree structures, add randomisation parameters for better learning, proportional shrinking of lead nodes on trees, and determines the depth of trees used as weak learners using a penalisation parameter added to prevent trees with high depth that prevents the model for overfitting and improves the performance. Categorical boosting (CatBoost) is a gradient boosting algorithm that successfully handles categorical features and numerical variables. CatBoost takes advantage of the categorical features by dealing with them during training instead of pre-processing time. The CatBoost algorithm reduces or avoids overfitting by using a new schema for calculating leaf values when selecting tree structures (Dorogush et al., 2018; Prokhorenkova et al., 2019).

### 2.2.3 Recursive and Deep Models

Non-recursive neural networks prediction accuracy is limited as no previous events are returned. Hochreiter & Schmidhuber (1997) introduced a deep neural network model known as long short-term memory (LSTM). The LSTM cell is trained to learn what can be forgotten, making it possible to keep information over a long period (Sautermeister, 2016). The LSTM uses three gates that can be trained to accumulate or remove information from its current state. One deep learning method that has become more commonly used in the solar field is CNN, especially for image classification problems (Edun et al., 2021; Gao et al., 2020). The CNN can have tens or hundreds of layers where each layer can learn to detect different features of an image. Filters are applied to each training image at different resolutions such as convolution, rectified linear unit, and pooling to learn specific data features. The convolutional layer is a tensor with the images' shape and input channels' shape. The result is set into feature map channels, and ReLu is used to allow faster training by maintaining positive values. The pooling layer simplifies the output by performing a nonlinear down-sampling that reduces the network's parameters that must be learned. A schematic diagram summarising the predictors and predicted parameters and the machine learning algorithms deployed in this study is given in Fig. 2.

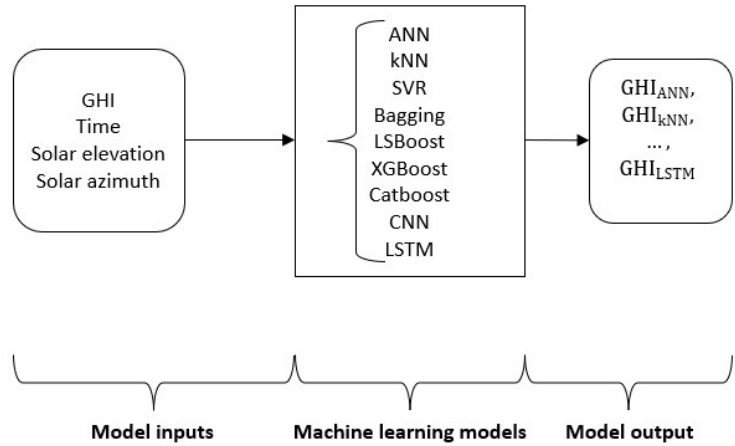

Fig. 2 Diagram for the site adaptation model. The inputs are GHI and time from STRÅNG, solar azimuth, and solar elevation for the specific location, while the output from the model is the weather station GHI.

### 2.3 Hyperparameter optimisation

The hyperparameters are parameter values in machine learning algorithms that is used to control the learning process, and these values cannot be changed during training. Therefore, hyperparameters must be tuned to attain the best accuracy. The optimal hyperparameter values vary depending on the problem, as the parameters are tuned with a specific dataset (Burkov, 2019). One typical way to tune these parameters is to experimentally find the best combination of values, and one of the most straightforward hyperparameter tuning strategies is grid search (Yu & Zhu, 2020). The grid search uses several given hyperparameter values and trains the model with all possible combinations from the given values. From the grid search, the model with the best performance will be kept. This study uses 20% of the training data for hyperparameter optimisation. In the sensitivity analysis, each training set has a hyperparameter optimisation. A grid search is performed to find the optimal hyperparameters. The chosen hyperparameters and range of grid searches optimised in this study are summarised in Table 3.

Table 3: Hyperparameters used for tuning with grid search.

| Estimation Models | Hyperparameter optimisation |
|---|---|
| SVR | Control error:[0.1,1,10,100,1000], Gamma:[1,0.1,0.01,0.001,0.0001] |
| k-NN | N_neighbors:[3,5,11,19] |
| Bagging | Learning_cycles:[10,20,40,80,100] |
| LSBoost | Max_depth: [2,3,...,9,10], Learning_cycles:[10,20,40,80,100] |
| XGBoost | Max_depth:[3,5,7,9], Learning_rate:[0.03,0.05,0.07,0.1], Learning_cycles:[15,30,50,100,200] |
| CatBoost | Max_depth:[3,5,7,9], Learning_rate:[0.03,0.05,0.07,0.1], Learning_cycles:[15,30,50,100,200] |
| ANN | Epochs:[5,20,40,100,200] |
| CNN | Batch_size:[20,40,60], Epochs:[50,100,150,200] |
| LSTM | Batch_size:[20,40,60], Epochs:[50,100,150,200], Hidden_layers:[5,10,25,50,100,200] |

### 2.4 Performance analysis

The error metrics used in this study to evaluate the performance of the machine learning algorithms for site adaption are the mean absolute error (MAE), mean bias error (MBE), and r-squared ($R^2$) with the following formulae:

$$\text{MAE} = \frac{1}{n}\sum_{i=1}^{n}|y_{\text{estimated}} - y_{\text{observed}}|, (1)$$

$$\text{MBE} = \frac{1}{n}\sum_{i=1}^{n}(y_{\text{estimated}} - y_{\text{observed}}), (2)$$

$$R^2 = 1 - \frac{\sum|y_{\text{estimated}} - y_{\text{observed}}|}{\sum|y_{\text{estimated}} - y_{\text{mean}}|}, (3)$$

The MAE and MBE should be zero if the estimated values are 100% accurate, while the $R^2$ should be one.

### 2.5 Sensitivity analysis

A sensitivity analysis is conducted in this study to understand how the STRÅNG quality varies depending on the chosen training period. The data between 2008–2013 is split into five different datasets where each specific dataset contains one year of training data and four years of testing data. Cross-validation is used to analyse the accuracy variations dependent on the training years' choice.

## 3 Results and discussions

This section provides the site-adaptation results of the machine-learning models for Kiruna, Norrköping, and Visby. The machine learning models are benchmarked against STRÅNG and linear regression, and quantile mapping, respectively. The linear regression and quantile mapping are used to see if machine learning models can perform better than common site adaptation techniques. The machine learning models could reduce the MAE, MBE and increase the $R^2$ significantly using one year of training data. In Fig. 3, the accuracy of the respective machine learning model is presented using training data from 2008–2009 and testing data from 2010–2015 in Kiruna.

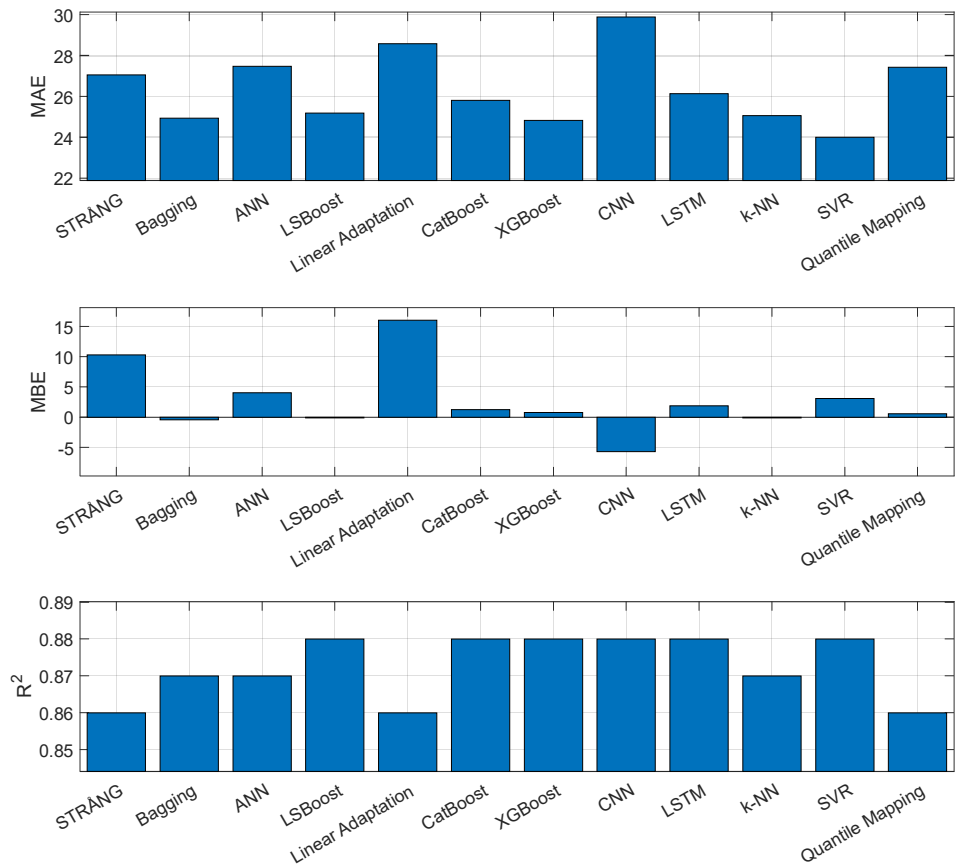

Fig. 3 Estimation error measures from 2010-2015 on the models in Kiruna.

In Table 4, the performance metrics, including MAE, MBE, and $R^2$ are mapped for all selected machine learning models and the traditional statistical approaches using training data from 2008–2009 and testing data from 2010–2015 in all three locations: Kiruna, Norrköping, and Visby.

Table 4: Estimation error measures of all machine learning models and traditional statistical approaches from 2010-2015 in Kiruna, Norrköping, and Visby.

| Location | Models | MAE | MBE | $R^2$ |
|---|---|---|---|---|
| Kiruna | STRÅNG | 27.05 | 10.32 | 0.86 |
| | Linear Adaptation | 28.57 | 16.01 | 0.86 |
| | Quantile Mapping | 27.41 | 0.59 | 0.86 |
| | Bagging | 24.93 | −0.42 | 0.87 |
| | k-NN | 25.05 | **−0.12** | 0.87 |
| | SVR | **24.00** | 3.09 | **0.88** |
| | LSBoost | 25.18 | −0.13 | **0.88** |
| | XGBoost | 24.82 | 0.78 | **0.88** |
| | CatBoost | 25.80 | 1.25 | **0.88** |
| | ANN | 27.47 | 4.03 | 0.87 |
| | CNN | 28.88 | −5.71 | **0.88** |
| | LSTM | 26.13 | 1.85 | **0.88** |
| Norrköping | STRÅNG | 31.37 | 6.46 | 0.89 |
| | Linear Adaptation | 31.57 | 8.89 | 0.89 |
| | Quantile Mapping | 32.86 | −5.25 | 0.89 |
| | Bagging | 32.47 | 2.66 | 0.88 |
| | k-NN | **31.05** | **−2.57** | 0.89 |
| | SVR | 31.56 | 6.94 | 0.89 |
| | LSBoost | 33.80 | 4.03 | 0.88 |
| | XGBoost | 37.13 | 9.93 | 0.83 |
| | CatBoost | 35.26 | 5.49 | 0.89 |
| | ANN | 34.37 | 4.76 | 0.88 |
| | CNN | 38.80 | 11.98 | **0.90** |
| | LSTM | 36.87 | 14.90 | **0.90** |
| Visby | STRÅNG | 31.29 | 7.39 | 0.90 |
| | Linear Adaptation | 31.38 | 8.10 | 0.90 |
| | Quantile Mapping | 32.89 | −4.66 | 0.90 |
| | Bagging | 33.46 | 5.36 | 0.90 |
| | k-NN | **30.92** | **−0.03** | 0.91 |
| | SVR | 32.99 | 9.87 | 0.90 |
| | LSBoost | 35.66 | 7.68 | 0.90 |
| | XGBoost | 40.18 | 14.60 | 0.84 |

| | | | |
|---|---|---|---|
| CatBoost | 37.08 | 9.77 | 0.90 |
| ANN | 33.56 | 2.09 | 0.90 |
| CNN | 35.59 | 7.78 | 0.91 |
| LSTM | 36.12 | 20.39 | **0.92** |

As presented in Fig. 3 the STRÅNG solar radiation data had a high variation in accuracy dependent on location. In Kiruna, the MBE is 10.32 (W/m$^2$) compared to 6.46 (W/m$^2$) and 7.39 (W/m$^2$) in Norrköping and Visby, respectively. Overall, machine learning models improved the solar radiation estimations made by STRÅNG as compared to traditional statistical approaches. However, the machine learning models' accuracy varied depending on location, and none of the models outperformed the rest of the models at the sites being studied. Therefore, the results show that there is no machine learning model that can be used as a universal model for site adaptation. The choice of machine learning model should be thoroughly analysed since the most accurate model depends on the variability of the meteorology at the given site. The deep learning models did not improve the MBE at all sites. More complex machine learning models increase the bias despite being more computationally expensive. It might be concluded that the dataset used for site adaptation could therefore be too small for deep learning models to find the complex relationships in the training dataset. The SVR model had promising results in cross-validation at all given locations. However, the SVR model had a longer computational time than the deep learning models. The time complexity of SVR is $O(N^3)$ (Hui et al., 2016). Therefore large sizes of training will be problematic for the SVR model. On the other hand, shallow machine learning models such as k-NN are easy to implement as a site adaptation technique with the ability to improve gridded solar products without being computational heavy. Nevertheless, overall, it could be seen that several of the chosen machine learning models had a significant reduction in bias compared to the STRÅNG model, which is necessary for decision-makers as it reduces the risk in projects and financial investments.

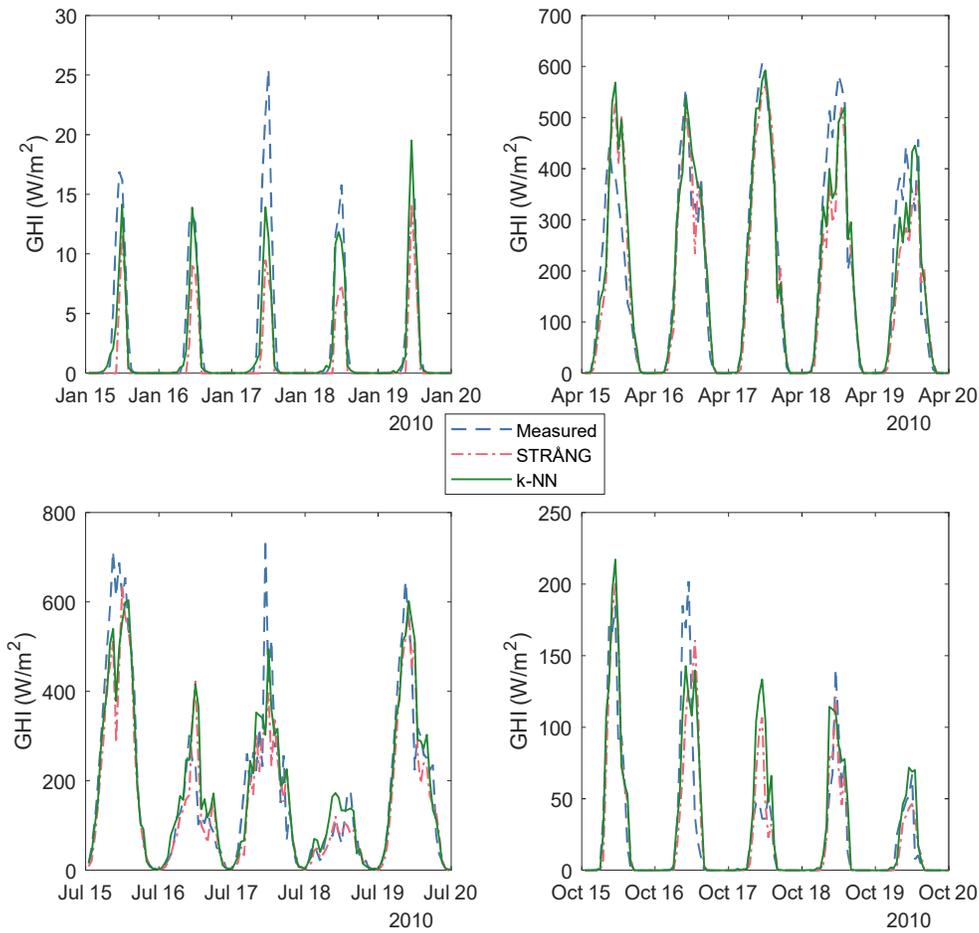

Fig. 4 Site adaptation with k-NN in Kiruna. The horizontal axis is time in days, and the vertical axis is GHI. The dashed line represents the weather station GHI, the dashed dot line represents the STRÅNG GHI, and the solid line represents the k-NN GHI.

Fig. 4 presents the result of five-day estimations obtained with the k-NN model in Kiruna for five days after the equinox of spring, after the solstice of summer, after the equinox of autumn, and after the solstice of winter, respectively. The days presented show the variation of GHI dependent on the meteorological season. The k-NN model performed well in all-weather seasons. However, in cases where STRÅNG estimates global horizontal irradiance poorly, similar estimates can be noted for k-NN. Therefore, it is essential to reduce the variance in the model's GHI estimations, and one solution would be to develop monthly trained models instead. However, monthly trained models require several models as one model cannot be used to estimate a whole year. In Fig. 5, the cross-validation is presented using k-NN with five years as testing data to analyse the accuracy of k-NN using different years of STRÅNG data as the quality can vary from year-to-year depending on meteorological conditions. The cross-validation results for all models can be found in the Appendix. The solar radiation estimation accuracy is reported in terms of $R^2$. MAE. and MBE for Kiruna, Norrköping, and Visby. The k-NN accuracy varies significantly using different years of STRÅNG data as input. In 2010 the MBE is −14.97 (W/m$^2$) in Norrköping but at Visby and Kiruna the MBE is −4.96 (W/m$^2$) and −2.60 (W/m$^2$), respectively. The significantly lower MBE at Visby and Kiruna in 2010 shows that the k-NN estimations using the STRÅNG data as the input varies at the different locations yearly, and it should be noticed that the yearly STRÅNG dataset used for training should be analysed for the specific location. The variability of solar radiation from STRÅNG from year to year at the different sites shows that both the machine learning models and the training dataset used for site adaptation will vary at any given location. The cross-validation shows that several of the chosen machine learning models can significantly reduce the estimation bias from STRÅNG at Kiruna, Norrköping, and Visby for the respective dataset.

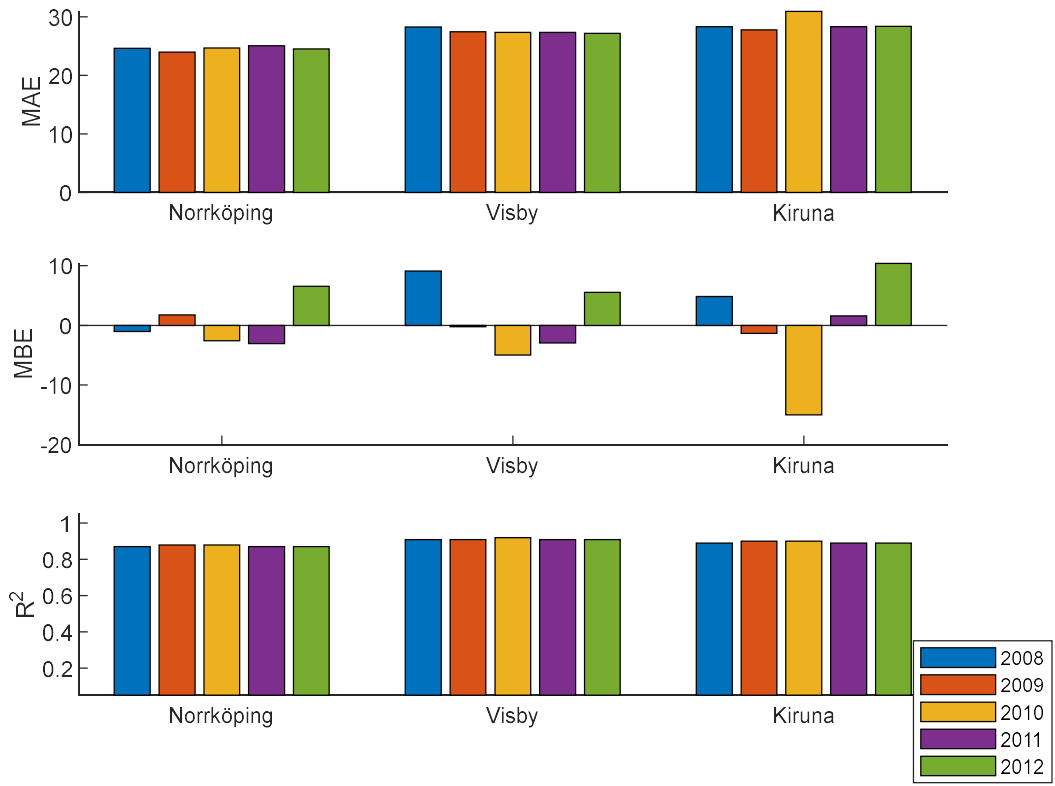

Fig. 5 Estimation error measures on k-NN in Kiruna, Norrköping, and Visby dependent on the training set.

# 4  Conclusion

In this study, machine learning, as a site-adaptation strategy, has improved the quality of global horizontal irradiance (GHI) estimates of STRÅNG. STRÅNG is a mesoscale model that produces GHI data at a temporal resolution of 1 h, covering the Nordic countries with a 2.5 km × 2.5 km spatial resolution. A total of nine machine-learning algorithms are considered: k-NN, SVR, Bagging, LSBoost, XGBoost, CatBoost, Artificial Neural Network (ANN), Convolutional Neural Network (CNN), and long short-term memory (LSTM), and all of which are benchmarked against the traditional linear mapping and quantile mapping, as well as against the raw STRÅNG product. In particular, two main conclusions can be drawn:

- Machine learning as a site-adaptation technique can substantially improve the STRÅNG model's quality. Indeed, STRÅNG data have shown improvements at all analysed locations, Kiruna, Norrköping, and Visby, after site adaptation.
- The inter-annual variability and spatial inhomogeneity in solar radiation affect the quality of the STRÅNG model. In that, no universal machine-learning model can be said to be optimal for every location.

## Acknowledgements

The authors acknowledge the following funding agencies and related projects for the development of machine learning algorithms for different energy systems applications: Vinnova for the project "SnowSat-An AI approach towards efficient hydropower production" and the Swedish Energy agency for the project "Evaluation of the first agrivoltaic system in Sweden" [grant number 51000-1].

## Appendix

The cross-validation error metrics MAE, $R^2$, and MBE for the selected location between 2008–2013 are summarised Table. A1, Table. A2, and Table. A3.

Table. A1. Cross-validation MAE measures in Kiruna, Norrköping, and Visby.

| Location | Models | MAE | | | | |
|---|---|---|---|---|---|---|
| | | 2008 | 2009 | 2010 | 2011 | 2012 |
| Kiruna | STRÅNG | 26.86 | 26.60 | 26.23 | 26.85 | 26.60 |
| | Linear Adaptation | 29.92 | 29.02 | 29.82 | 29.16 | 28.43 |
| | Quantile Mapping | 27.71 | 27.50 | 27.54 | 29.27 | 27.32 |
| | Bagging | 24.57 | 23.78 | 24.52 | 24.19 | 24.11 |
| | k-NN | 24.64 | 23.98 | 24.67 | 25.06 | 24.54 |
| | SVR | **23.75** | **22.49** | **22.67** | **22.76** | **23.87** |
| | LSBoost | 24.87 | 24.24 | 24.60 | 24.48 | 24.44 |
| | XGBoost | 24.33 | 23.61 | 24.10 | 23.89 | 24.09 |
| | CatBoost | 25.46 | 23.66 | 24.29 | 23.82 | 24.09 |
| | ANN | 29.21 | 29.13 | 27.01 | 29.11 | 27.16 |
| | CNN | 25.71 | 25.54 | 24.60 | 24.70 | 24.91 |
| | LSTM | 25.58 | 23.86 | 25.95 | 23.38 | 26.91 |
| Norrköping | STRÅNG | 32.35 | 31.53 | 29.91 | 31.81 | 31.60 |
| | Linear Adaptation | 34.08 | 34.41 | 37.77 | 34.35 | 33.53 |
| | Quantile Mapping | 32.71 | 33.51 | 39.63 | 32.45 | 31.95 |
| | Bagging | 28.02 | 26.78 | **27.89** | 26.93 | 27.21 |
| | k-NN | 28.32 | 27.75 | 30.93 | 28.34 | 28.35 |
| | SVR | 27.62 | **26.03** | 28.39 | **25.82** | **26.07** |
| | LSBoost | 28.52 | 27.61 | 28.58 | 27.93 | 28.08 |
| | XGBoost | **27.50** | 26.93 | 27.95 | 26.84 | 27.27 |
| | CatBoost | 29.24 | 27.26 | 28.24 | **25.82** | 27.80 |
| | ANN | 35.15 | 30.98 | 36.33 | 34.35 | 33.27 |
| | CNN | 27.87 | 27.81 | 28.93 | 28.85 | 30.79 |
| | LSTM | 28.58 | 28.52 | 27.93 | 27.01 | 26.22 |
| Visby | STRÅNG | 32.42 | 31.79 | 31.01 | 31.51 | 31.88 |
| | Linear Adaptation | 33.96 | 35.15 | 35.56 | 35.81 | 33.22 |
| | Quantile Mapping | 32.59 | 32.64 | 35.79 | 33.38 | 32.93 |
| | Bagging | 27.38 | 26.81 | 25.91 | 26.28 | 26.23 |
| | k-NN | 28.24 | 27.46 | 27.35 | 27.32 | 27.18 |
| | SVR | **25.10** | **24.97** | **24.50** | **24.84** | **24.50** |
| | LSBoost | 28.25 | 27.04 | 26.41 | 27.12 | 27.42 |
| | XGBoost | 27.16 | 26.26 | 25.45 | 25.94 | 26.40 |
| | CatBoost | 27.30 | 26.26 | 25.75 | 26.20 | 26.55 |
| | ANN | 34.43 | 35.07 | 30.79 | 31.51 | 32.60 |
| | CNN | 27.38 | 29.45 | 28.74 | 28.69 | 27.51 |
| | LSTM | 26.39 | 27.65 | 25.57 | 31.65 | 27.56 |

Table. A2. Cross-validation $R^2$ measures in Kiruna, Norrköping, and Visby.

| Location | Models | $R^2$ | | | | |
|---|---|---|---|---|---|---|
| | | 2008 | 2009 | 2010 | 2011 | 2012 |
| Kiruna | STRÅNG | 0.86 | 0.86 | 0.87 | 0.86 | 0.87 |
| | Linear Adaptation | 0.86 | 0.87 | 0.88 | 0.87 | 0.87 |
| | Quantile Mapping | 0.86 | 0.86 | 0.87 | 0.86 | 0.86 |
| | Bagging | **0.88** | 0.88 | 0.88 | 0.88 | 0.87 |
| | k-NN | 0.87 | 0.88 | 0.88 | 0.87 | 0.87 |
| | SVR | **0.88** | **0.89** | **0.89** | **0.89** | 0.87 |
| | LSBoost | **0.88** | **0.89** | **0.89** | 0.88 | 0.88 |
| | XGBoost | **0.88** | **0.89** | **0.89** | 0.88 | 0.88 |
| | CatBoost | **0.88** | **0.89** | **0.89** | **0.89** | **0.89** |
| | ANN | 0.87 | 0.87 | 0.88 | 0.87 | 0.87 |
| | CNN | 0.87 | 0.88 | 0.88 | 0.88 | 0.87 |
| | LSTM | **0.88** | **0.89** | **0.89** | **0.89** | 0.88 |
| Norrköping | STRÅNG | 0.88 | 0.88 | 0.90 | 0.88 | 0.89 |
| | Linear Adaptation | 0.90 | 0.90 | 0.91 | 0.91 | 0.90 |
| | Quantile Mapping | 0.90 | 0.90 | 0.91 | 0.90 | 0.90 |
| | Bagging | 0.91 | 0.91 | 0.92 | **0.92** | 0.92 |
| | k-NN | 0.91 | 0.91 | 0.92 | 0.91 | 0.91 |
| | SVR | **0.92** | **0.92** | **0.93** | **0.92** | 0.92 |
| | LSBoost | **0.92** | **0.92** | **0.93** | **0.92** | 0.92 |
| | XGBoost | **0.92** | **0.92** | **0.93** | **0.92** | 0.92 |
| | CatBoost | **0.92** | **0.92** | **0.93** | **0.92** | **0.93** |
| | ANN | 0.90 | 0.90 | 0.92 | 0.91 | 0.91 |
| | CNN | **0.92** | **0.92** | 0.92 | **0.92** | 0.92 |
| | LSTM | **0.92** | **0.92** | **0.93** | **0.92** | 0.92 |
| Visby | STRÅNG | 0.90 | 0.90 | 0.91 | 0.90 | 0.90 |
| | Linear Adaptation | 0.90 | 0.90 | 0.91 | 0.91 | 0.90 |
| | Quantile Mapping | 0.90 | 0.90 | 0.91 | 0.90 | 0.90 |
| | Bagging | 0.91 | 0.91 | 0.92 | **0.92** | 0.92 |
| | k-NN | 0.91 | 0.91 | 0.92 | 0.91 | 0.91 |
| | SVR | **0.92** | **0.92** | **0.93** | **0.92** | 0.92 |
| | LSBoost | **0.92** | **0.92** | **0.93** | **0.92** | 0.92 |
| | XGBoost | **0.92** | **0.92** | **0.93** | **0.92** | 0.92 |
| | CatBoost | **0.92** | **0.92** | **0.93** | **0.92** | **0.93** |

| | | | | | |
|---|---|---|---|---|---|
| ANN | 0.90 | 0.90 | 0.92 | 0.91 | 0.91 |
| CNN | **0.92** | **0.92** | 0.92 | **0.92** | 0.92 |
| LSTM | **0.92** | **0.92** | 0.93 | **0.92** | 0.92 |

Table. A3. Cross-validation MBE measures in Kiruna, Norrköping, and Visby.

| Location | Models | MBE | | | | |
|---|---|---|---|---|---|---|
| | | 2008 | 2009 | 2010 | 2011 | 2012 |
| Kiruna | STRÅNG | 8.68 | 9.07 | 8.06 | 7.95 | 10.11 |
| | Linear Adaptation | −2.23 | −0.86 | −4.66 | −5.74 | 5.59 |
| | Quantile Mapping | −1.23 | 0.65 | −4.53 | −6.19 | 5.89 |
| | Bagging | −1.51 | 0.57 | −3.07 | −3.39 | 5.52 |
| | k-NN | −1.05 | 1.73 | −2.60 | −3.04 | 6.51 |
| | SVR | 2.20 | 1.12 | −1.80 | **−1.70** | 7.31 |
| | LSBoost | −0.61 | 0.59 | −2.79 | −3.06 | 6.03 |
| | XGBoost | **0.11** | 1.29 | −2.38 | −2.53 | 6.68 |
| | CatBoost | 0.66 | **0.40** | −2.73 | −2.54 | 6.09 |
| | ANN | 1.94 | 0.81 | **0.39** | −4.48 | 7.69 |
| | CNN | 0.96 | −2.71 | −1.69 | −3.31 | **3.63** |
| | LSTM | −2.16 | 3.57 | −4.36 | −2.13 | 2.48 |
| Norrköping | STRÅNG | 10.66 | 8.92 | 6.26 | 9.47 | 11.42 |
| | Linear Adaptation | 3.98 | −4.14 | −19.66 | −0.69 | 9.77 |
| | Quantile Mapping | 5.58 | −4.16 | −20.49 | −0.89 | 9.38 |
| | Bagging | 5.37 | −2.32 | −11.90 | **−0.30** | 8.83 |
| | k-NN | 4.85 | −1.36 | −14.97 | 1.59 | 10.35 |
| | SVR | 6.09 | −2.97 | −12.96 | −0.84 | 9.04 |
| | LSBoost | **−2.40** | −12.61 | **−0.09** | 9.22 | 7.58 |
| | XGBoost | 6.06 | −1.99 | −11.17 | 0.49 | 9.98 |
| | CatBoost | 7.22 | −2.07 | −10.60 | −0.84 | 9.12 |
| | ANN | 8.43 | **0.33** | −16.73 | 3.04 | 15.11 |
| | CNN | 3.19 | −3.69 | −9.78 | −8.07 | **−1.11** |
| | LSTM | 11.41 | −4.14 | −8.50 | 6.50 | 8.32 |
| Visby | STRÅNG | 11.41 | 9.52 | 8.34 | 8.67 | 10.68 |
| | Linear Adaptation | 7.65 | −3.18 | −8.65 | −7.27 | 3.38 |
| | Quantile Mapping | 7.73 | −1.73 | −11.45 | −5.97 | 4.26 |

| | | | | | |
|---|---|---|---|---|---|
| Bagging | 7.00 | −1.99 | −4.47 | −5.19 | 4.43 |
| k-NN | 9.10 | **−0.20** | −4.96 | −2.96 | 5.51 |
| SVR | **5.29** | −4.82 | −5.35 | −6.58 | 3.41 |
| LSBoost | 7.58 | −2.61 | −4.51 | −5.08 | 4.73 |
| XGBoost | 8.48 | −2.01 | −3.68 | −4.36 | 5.27 |
| CatBoost | 7.29 | −2.29 | **−3.34** | −4.87 | 5.16 |
| ANN | 5.36 | 2.94 | −5.03 | **−2.54** | 8.25 |
| CNN | 8.72 | −7.84 | −8.04 | −6.92 | **0** |
| LSTM | 8.32 | −4.15 | **−3.34** | −16.90 | 2.87 |